\documentclass{ws-procs961x669}            % default, citations in superscript

\begin{document}
\title{Studying the EPRL spinfoam self-energy}

\author{Pietropaolo Frisoni}

\address{Department of Physics and Astronomy, Western University,\\
London, Ontario, Canada\\
E-mail: pfrisoni@uwo.ca\\
}

\begin{abstract}
I present some recent progresses in the study of the EPRL self-energy amplitude \cite{MY_PAPER}. New numerical methods allow to analyze how the divergence scales, for which previous works provided very different lower and upper bounds. I show the role that the Immirzi parameter plays in the asymptotic behavior, and I discuss the dynamical expectation value of the dihedral angle boundary observable.
\end{abstract}

\bodymatter

\section{Introduction}
\label{sec:introduction}
The spinfoam formalism is an attempt to define the dynamics of loop quantum gravity in an explicitly Lorentz covariant way \cite{Perez:2012wv, Rovelli2004}. It defines transition amplitudes for spinnetwork states of the canonical theory in a form of a sum (or equivalently a refinement \cite{Rovelli:2010qx}) over all the possible two-complexes having the chosen spinnetwork as boundary. This is equivalent to a sum over histories of quantum geometries providing in this way a regularised version of the quantum gravity path integral. \\ \\
The state of the art of the spinfoam approach to LQG is currently the model proposed by Engle, Pereira, Rovelli and Livine (EPRL) \cite{Pereira:2007nh, Engle2007}, independently developed by Freidel and Krasnov \cite{Freidel:2008fk} and extended to arbitrary spinnetwork states \cite{Ding:2010fw, Kaminski:2009fm}. I assume that the reader is familiar with the EPRL-FK\footnote{from now on I will call it just EPRL for notation convenience.} model, and refer to the original literature \cite{Engle:2007wy, Livine:2007vk, Livine:2007ya, Freidel:2007py} and existing reviews (e.g. \cite{Perez:2012wv, book:Rovelli_Vidotto_CLQG, Rovelli2011c, Dona:2010cr}) for the details on how it encodes the covariant dynamics of Loop Quantum Gravity. \\ \\
The EPRL model presents infrared divergences, which have a similar structure to the UV divergences in the Feynman expansion of a standard quantum field theory. The presence of divergences require a renormalization procedure. This is an important open direction of investigation in the theory, since their study and understanding is important in the definition of the continuum limit. A number of questions are still open, for example regarding the proper normalisation of the $n$ point functions and similar \cite{book:Rovelli_Vidotto_CLQG}. Divergences have been the subject of many studies following different investigation strategies: via refining of the 2-complex as proposed in \cite{dittrich_2016_continuum_limit, Dittrich_decorated_tensor_network, Bahr_Investigation_of_the_spinfoam}, or via a resummation, defined for instance using group field theory/random tensor models as proposed in \cite{Bonzom_random_tensor_models, Benedetti_phase_transition_in_dually, Carrozza_renormalization_of, Geloun_functional_renormalization}. The properties of these divergences have been studied in the context of the Ponzano-Regge model of 3d quantum gravity and discrete BF theory \cite{Bonzom:2010zh, Bonzom:2011br}, group field theory \cite{Baratin_melonic_Phase_Transition} and EPRL model: with both Euclidean \cite{Perini:2008pd, Krajewski:2010yq} and Lorentzian signature \cite{Riello:2013bzw}.  \\ \\
Analytic estimates of divergences in the Lorentzian EPRL model \cite{Riello:2013bzw} and \cite{Dona:2018infrared} are based on two different methods. In \cite{Riello:2013bzw} the ``self-energy" amplitude is considered, finding a logarithmic divergence $\log{K}$ as a \emph{lower bound}, where $K$ is an artificial cut-off on the internal $SU(2)$ spins associated with the faces of the diagrams (the need to introduce a cut-off is justified by the fact that the sum on the internal spins is unbounded). The computation is rather involved and relies on the techniques developed for the asymptotic analysis of the vertex amplitude of the model \cite{article_Barrett_etal_2010_lorentzian_spinfoam_amplitude}. This approach requires an independent study of each geometrical sector and the logarithmic divergence is obtained by looking at the non-degenerate geometries. On the other hand, in \cite{Dona:2018infrared} is proposed an algorithm to systematically determine the potential infrared divergence for all spinfoam diagrams. The approach is based on the hypothesis that the dominant contribution to the divergence scaling of the amplitude comes from the uniform scaling of all the spins and that there is no interference between various terms of the sum. This leads to an \textit{upper bound} on the divergence proportional to $K^9$. \\ \\
These two bounds are therefore remarkably different, leading a window of possibilities which spans several powers in the cutoff, and it is necessary to find methods of investigation that can help clarify the question. Here I report the results of a numerical study \cite{MY_PAPER} regarding the scaling of the self-energy amplitude, which also investigates the dependence on the Immirzi parameter and some other boundary data. In particular, the numerical estimates confirm both the two bounds previously found. \\  \\
For the numerical calculation of the EPRL vertex amplitudes, the ``sl2cfoam-next" library was used \cite{Francesco_draft_new_code}. All computations were performed on Compute Canada's Cedar and Graham clusters (www.computecanada.ca). The plots were made with julia \cite{bezanson2017julia} and Mathematica \cite{Mathematica}. Here I present a summary of the results obtained. The interested reader can find further details and analysis, including explicit formulas for the $SU(2)$ invariant symbols and the booster functions, in the full analysis \cite{MY_PAPER}. \\ \\
The paper is organized as follows. In section \ref{sec:self_energy_diagram} I describe the triangulation used to study the self-energy amplitude. In section \ref{sec:derivation_BF_amp} the amplitude is derived in the purely $SU(2)$ BF model. This intermediate step allows to compare the numerical results with the analytical ones present in the literature. In section \ref{sec:From_BF_to_EPRL} the EPRL amplitude is derived from the BF model. In section \ref{sec:Div_analysis} I perform the numerical analysis of the EPRL self-energy amplitude. Finally, in section \ref{sec:Boundary_observables}, I report the results of the dynamical expectation value of the boundary angle operator.

\section{Self-energy diagram} 
\label{sec:self_energy_diagram}
\begin{figure}[tb]
    \centering
    \includegraphics[width=4.5cm]{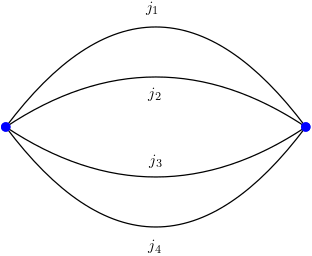}
    \caption{\emph{Boundary graph of the spinfoam associated with the self-energy amplitude. In blue are highlighted the two four valent nodes, corresponding to the two tetrahedra in the dual triangulation. The nodes are joined by all the links. The boundary spins are denoted with $j_{f}$, where $f = 1, . . . , 4$}} 
    \label{fig:bubblediagram}
\end{figure}

I consider bubbles, studying one of the most elementary diagrams appearing in the self-energy amplitude \cite{Riello:2013bzw, Dona:2018infrared}. Since the associated divergence can be viewed as the one related to a particularly simple triangulation, from a computational point of view it turns out to be the one of the simplest divergence to deal with. The triangulation is formed by two 4-simplices joined by four tetrahedra. The associated two-complex turns out to be composed by two vertices, four edges, six internal faces (one per couple of edges) and four external faces (one per edge). The boundary of the dual triangulation is formed by two tetrahedra joined by all the faces, therefore the boundary graph consists in two four valent nodes connected by all the links (see Fig. \ref{fig:bubblediagram}). \\
The kinematical Hilbert space of LQG at fixed graph $\Gamma$, with $L$ links and $N$ nodes, is:
\begin{equation}
\label{eq:Hilbert_space}
\mathcal{H}_{\Gamma} = L_2 \left[ SU(2)^L / SU(2)^N \right]
\end{equation}
For the self-energy diagram, we have $L = 4$, $N = 2$ and the corresponding graph\footnote{from now on, I will drop the explicit dependence on $\Gamma$} $\Gamma$ is represented in Figure \ref{fig:bubblediagram}. 
Since the self-energy spinfoam graph is entirely symmetrical and has two nodes, I shall use the $+$ and $-$ symbols as labels for the latter. I use the same symbols to distinguish the corresponding intertwiners $i_{\pm}$, i.e. the invariant $SU(2)$ tensors associated with each node. 
The intertwiner space of a 4-valent node is denoted as:
\begin{equation}
\label{eq:intertwier_space}
\mathcal{I}_4 = Inv \left[ V^{j_1} \otimes V^{j_2} \otimes V^{j_3} \otimes V^{j_4} \right]   
\end{equation}
where $V^{j_i}$ is the irreducible representation of spin $j_i$. I consider the recoupling base $(j_1, j_2)$ for the intertwiner space \eqref{eq:intertwier_space}. That is, I fix a pairing of the links at each node and I choose the basis that diagonalises the modulus square of the sum of the $SU(2)$ generators in the pair $(j_1, j_2)$. A basis for the Hilbert space \eqref{eq:Hilbert_space} is given by the spin-network states $|j_l,i_{\pm} \rangle$, where the $j_l$'s are the spin associated with each link of the graph, $l = 1...4$, and the $i_{\pm}$'s are a basis in the corresponding intertwiners space \eqref{eq:intertwier_space} according to the recoupling scheme. The spin-network states are interpreted as quantum tetrahedra or, if the node have valence higher than 4, as quantum polyhedra\footnote{at least to some extent, since the Heisenberg uncertainty principle prevent us from determining the full geometry of the polyhedra. Thus, the interpretation of a spinnetwork state as a sharp polyhedron fails} \cite{Bianchi:2011hp, Bianchi:2011ub}. The corresponding discrete geometry, in which the shared faces between the tetrahedra must have the same area but nor necessarily the same shape or orientation, is called \textit{twisted geometry} \cite{Freidel:2010aq}. \\
I shall focus on the subspaces $\mathcal{H}_j$ of \eqref{eq:Hilbert_space} such that all spins are the same $j_l=j$, for which the basis states are denoted as
\begin{equation}
  |j, i_{\pm}\rangle \equiv {\bigotimes}_{\pm}\ |i_{\pm}\rangle =  |i_+ \rangle \otimes  |i_{-}\rangle
\end{equation}
The explicit dependence on $j$ can be dropped in the notation, since all the spins have the same value.
\section{Derivation of the BF spinfoam amplitude} 
\label{sec:derivation_BF_amp}
I start deriving the formula for the spinfoam transition amplitude associated to the triangulation described in section \ref{sec:self_energy_diagram}. The strategy chosen to derive the self-energy EPRL amplitude consists in starting from the corresponding amplitude for the BF model, for which it is possible to compare our numerical computations with the exact estimate of the amplitude's divergence \cite{Dona:2018infrared}, since in the pure $SU(2)$ model one can isolate the divergent factor. Then, I include the contribution of the booster functions (see \cite{Speziale2016}), as they encode all the details of the EPRL model. This process serves to facilitate the derivation and, at the same time, make sure that the phases and normalization factors are correct. 
\subsection{Numerical scaling of the BF amplitude}
\label{subsec:numerical_scaling_BF}
\begin{figure}[tb]
    \centering
    \includegraphics[width=6.5cm]{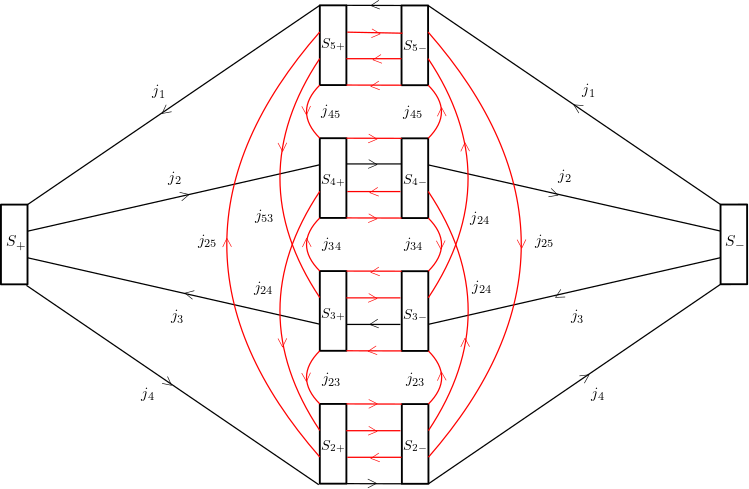}
    \caption{Spinfoam corresponding to the BF self-energy amplitude. The two $\{15j\}$ symbols are graphically represented in order to highlight the bulk contractions.} 
    \label{fig:15jsquare}
\end{figure}

\begin{figure}[tb]
    \centering
    \includegraphics[width=6cm]{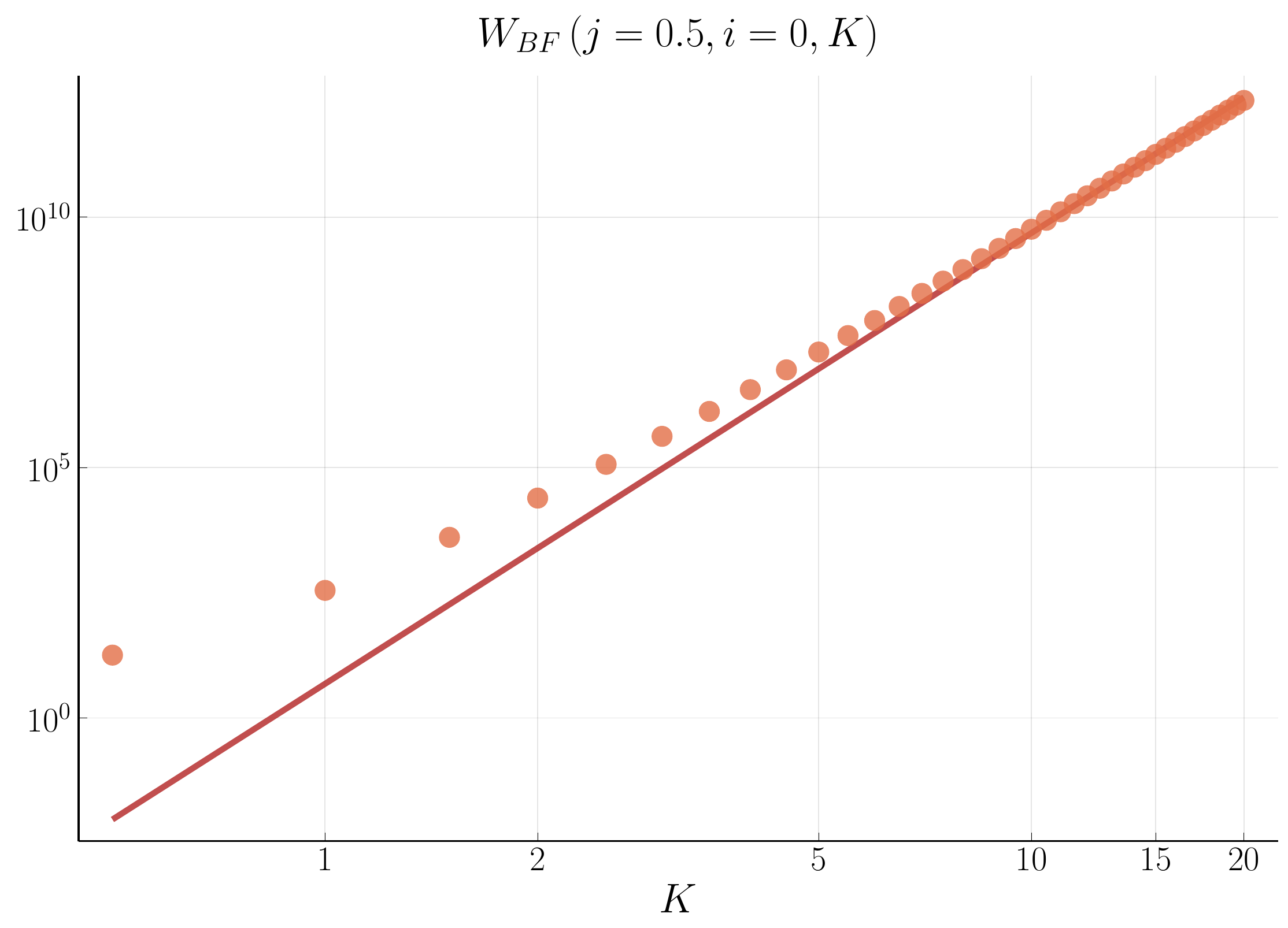}
    \caption{\emph{Log-log plot of the BF self-energy amplitude scaling. The points computed with \eqref{eq:BF_intertwiners_amplitude} are fitted with the curve $W_{BF} = 4.8 \cdot K^{9}$, and the latter is shifted so that the last point coincides with the end of the curve. Notice that the divergence scaling is reached for very low values of the bulk spins cutoff $K$.}} 
    \label{fig:bf_scaling}
\end{figure}

The BF self-energy amplitude is written as:
\begin{align}
W_{BF} \left( j, S_{+}, S_{-} \right) = &   \sum\limits_{j_{ab}}\prod\limits_{(a,b)}(2 j_{ab} +1) \hspace{1mm} \{15j \}^2
\label{eq:BF_simplicial_amplitude}
\end{align} 
where the algebric notation $\{15j \}^2$ refers to the spinfoam in Figure \ref{fig:15jsquare}. For the algebraic expression of the $SU(2)$ $\{ 15j\}$ invariant quantities, I refer to \cite{MY_PAPER, GraphMethods}.
In \eqref{eq:BF_simplicial_amplitude}, $(a,b) = (23,24,25,35,45,34)$ are the labels which denote the spins attached to the $6$ internal faces of the spinfoam, while the dependence on the $4$ boundary spins is denoted with $j \equiv j_f$ with $f = 1...4$. In \eqref{eq:BF_simplicial_amplitude}, the parameters specifying the boundary tetrahedra are generically denoted with $S$. The sums over the internal faces (explicitly highlighted in red) are unbounded: this is precisely the way in which the bubble's divergence manifests itself. Strictly speaking, the dimensional factor of the boundary faces should be taken into account in the expression for the amplitude, but this is a constant overall factor that can be ignored in the analysis, since it does not affect the scaling of the divergence. The 4-simplices are labeled in the following way. The subscripts $+$ and $-$ distinguish the tetrahedra belonging to one vertex amplitude from those of the other, while the numeric subscripts distinguish the ones belonging to the same vertex. The two boundary tetrahedra, on the ``opposite sides" of the spinfoam, are only labeled with $+$ and $-$ subscripts according the notation introduced in \ref{sec:self_energy_diagram}. The internal faces, dual to triangles, are labeled by two points indicating the tetrahedra attached to them, while the spins attached to the boundary links are labeled according to Figure \ref{fig:bubblediagram}. \\
In order to study numerically the scaling behavior of the amplitude, I impose an arbitrary cutoff $K$ on the spin's values of the $6$ internal faces. This allows to estimate the divergence scaling without performing an infinite number of sums. By doing so, the amplitude acquires an \textit{artificial} dependence also on $K$. From now on, I refer to $K$ as the ``bulk spins cut-off". \\
I perform the integration over $SU(2)$ and obtain:
\begin{align}
W_{BF} \left( j, i; K \right) = & (-1)^{2(i_+ + i_{-})} \sum\limits_{j_{ab}}\prod\limits_{(a,b)}(2 j_{ab} +1) \sum\limits_{i_e}\prod\limits_{e}(2i_{e} +1) \{ 15j\}^2 
\label{eq:BF_intertwiners_amplitude}
\end{align} 
I denoted with $i \equiv \left( i_{+} , i_{-} \right) $ the dependence on the intertwiners at the two nodes in the recoupling base $(j_1, j_2)$, and the phase $(-1)^{2(i_+ + i_{-})}$ comes from the fact that I changed the orientation of the line represented by the boundary intertwiners, according to our convention for the $\{15j\}$ symbol, following the rules of the $SU(2)$ graphical calculus \cite{book:varshalovic}. \\
In passing from \eqref{eq:BF_simplicial_amplitude} to \eqref{eq:BF_intertwiners_amplitude}, I attached an intertwiner $i_e$ with $e = 2, . . . , 5$ to each edge. Therefore, each edge carries a boundary spin, three face spins and an intertwiner. Triangular inequalities constrain the intertwiner to assume values in an interval centered on a face spin, implying that, for a fixed value of the spins on the faces, the sums over these intertwiners are bounded. I choose the spinnetwork boundary state with fixed spins and intertwiners and analyze the bubble's divergence with it, since the contraction of the amplitude with less trivial boundary states (for example, coherent states) requires to compute all the possible amplitudes obtained by varying independently the intertwinwers $i_{+} , i_{-}$, thus increasing the computational complexity. The computation with other choices of boundary states are considered in \cite{MY_PAPER}.  \\
As previously stated, I only consider equilateral boundary spins configuration, that is, the spins $j \equiv j_f = j_1 ... j_4$ have all the same value\footnote{notice that, by doing so, each boundary intertwiner assumes a range of possible integer values ranging from 0 to $2j+1$, so the overall phase of the amplitude is always unitary}. In fixing the boundary state entirely, that is, both the boundary spins and the intertwiners, the amplitude turns out to depend only on $6$ parameters, namely the $4$ spins $j \equiv j_f$ and the $2$ intertwinwers $i \equiv (i_{+}, i_{-})$. To further minimize the computational time, I fix the spins of the boundary faces to their minimum non-trivial value $j = \frac{1}{2}$. Notice that the value of the boundary spins $j$ and intertwiners $i$ is not relevant, since in the BF model the amplitude must be independent of the values of the latter. Computing the amplitude \eqref{eq:BF_intertwiners_amplitude} numerically, for different values of $K$, I verify that it correctly reproduces the divergent scaling derived in \cite{Dona:2018infrared}, as shown in Figure \ref{fig:bf_scaling}:
\begin{equation}
W_{BF} \left( j, i , K \right) \propto K^9.
\end{equation}
\section{From BF to EPRL}
\label{sec:From_BF_to_EPRL}
In the BF model it is possible to isolate the divergent factor, but the aim is to use the numerical tools to study the EPRL model, where only analytical estimates of the lower \cite{Riello:2013bzw} and upper bounds \cite{Dona:2018infrared} on the scaling of the bubble's divergence are available in the literature. The picture is therefore not clear, especially considering that these two limits differ greatly. In fact, while the lower bound is logarithmically divergent in the cutoff $K$ over the $SU(2)$ representation spins, the upper bound turns out to be $K^9$. \\
The crucial point of the passage from BF to EPRL model lies in the expression of the EPRL vertex amplitude. Since the reader may be unfamiliar with such amplitude form, I briefly underline the essential ideas, referring to the original paper  \cite{Speziale2016} for an accurate description.
\subsection{EPRL vertex amplitude} 
\label{subsec:EPR_vertex_amplitude}
The EPRL vertex amplitude is built from the topological $SL(2,\mathbb{C})$ spinfoam vertex amplitude by imposing, weakly, the simplicity constraints. This results in a restriction of the unitary irreducible representations in the principal series \cite{Engle:2007wy,Engle:2007uq}. In order to evaluate it in its original form, one should perform four group integrals (one of the original five integrals is redundant and has to be removed to guarantee finiteness \cite{Engle:2008ev}). Each group integral is, in general, a six dimensional unbounded highly oscillating integral for which numerical integration methods are not easy to implement efficiently. To get around this computational hurdle, an alternative form for the amplitude has been derived, resulting in a superposition of $SU(2)$ $\{15j\}$ symbols weighted by one booster functions $B4$ per edge in the considered vertex. 
\begin{align}
\label{eq:vertex_amplitude}
A_v \left(j_f, \, i_e , \gamma \right) &= \lim_{\Delta l \to \infty} \sum\limits_{l_{f} = j_f}^{j_f + \Delta l} \sum\limits_{k_{e}} \prod_{e} \left(2k_{e}+1\right) B_{4}(j_{f},l_{f};i_{e}, k_{e}; \gamma) \{15j\}_{j_f,i_1}(l_{f}, k_{e}) 
\end{align}
For the explicit formulas of the booster functions $B_4$, I refer to \cite{MY_PAPER, Speziale2016, Dona2019}. In \cite{MY_PAPER} it is also introduced a convenient graphical notation which combines the $SU(2)$ graphical calculus with algebraic formulas to define the quantities involved. Notice that in \eqref{eq:vertex_amplitude} the dimensional factors attached to the boundary intertwiners $i_e$ and to the boundary spins $j_f$ are neglected, since I assume the latter to be determined by the boundary spinnetwork state, so that they turn out to be constant. In less trivial spinfoams, as in the self-energy case, there are typically more vertex amplitudes glued together. This implies that dimensional factors of bulk spins and intertwiners must be taken into account, as did for the BF self-energy amplitude \eqref{eq:BF_intertwiners_amplitude}, in which the elementary vertex amplitude is represented by a single $SU(2)$ invariant $\{ 15j \}$ symbol. The key conceptual step to obtain the expression \eqref{eq:vertex_amplitude}, starting from the integral representation over $SL(2,C)$, is to decompose each $SL(2,C)$ integral $h$ according to the Cartan decomposition:
\begin{equation}
\label{eq:Cartan_decomp}
 h = u e^{(\frac{r \sigma_3}{2})} v^{-1},   
\end{equation}
where $u$ and $v$ are $SU(2)$ arbitrary rotations and $r\in[0,\infty)$ is the rapidity parameter of a boost along the $z$ axis. The compact integrals, resulting from the above parameterization, are then evaluated exactly composing the $SU(2)$ invariants in the amplitude \eqref{eq:vertex_amplitude}.
This decomposition results in a summation over a set of auxiliary spins $l_f$ for each face involving the vertex (excluding the gauge fixed one), for a total of $6$ distinct\footnote{the appearance of this infinite sum is actually linked to the splitting of the representations of the Lorentz group within the vertex and therefore exults from the Cartan decomposition \eqref{eq:Cartan_decomp}} $l_f$, with lower bound $l_f \geq j_f$, and a set of auxiliary intertwiners $k_e$ for each edge in the vertex (excluding the gauge fixed one) for a total of $4$, which can assume all the values compatible with triangular inequalities. Finally, the ``Y-map" imposes that the polyhedron shared by two adjacent polytopes lives in the same space-like hyperplane \cite{Dona:2010cr, article:bianchi_dona_speziale_2011_polyhedra_in_LQG, article:Dona_etal_2018_SU2_graph_invariants}. \\ \\
The Y-map is present only on one side of each booster function, namely the one reaching out to the next vertex: the group elements joining at the vertex are instead multiplied together without the latter. Because of this, an infinite sum appears on the auxiliary spins $l_f$ and the EPRL model is, in principle, recovered only in the limit in which this sum becomes infinite. The full EPRL amplitude is well defined \cite{Engle:2008ev}, as the summations over the $l_f$ are convergent. Nevertheless, in order to perform a numerical evaluation of the amplitude, we need to introduce a homogeneous cut-off\footnote{the nature of this cut-off is obviously completely different from the $K$ cutoff on the internal faces of self-energy amplitude introduced in section \ref{subsec:numerical_scaling_BF} for the $BF$ amplitude} $\Delta l$ on the auxiliary spins $l_f$. In the following, I will refer to the cutoff $\Delta l$ as the number of \textit{shell}. \\ \\
Beside the numerical precision with which the single terms that contribute to the vertex amplitude are computed, notice that this is the only approximation on which this method is based. If the convergence in the sum over the auxiliary spins $l_f$ is sufficiently good, then we obtain a reasonable estimate of the EPRL model. As originally discussed in \cite{Speziale2016}, the largest contributions to the booster functions should come from configurations with $l_f = j_f$, namely the minimal admissible values for the $l_f$ spins. The EPRL model defined with the approximation $l_f = j_f$ is usually called ``simplified model". Even if the convergence of the amplitude \eqref{eq:vertex_amplitude} as a function of $\Delta l$ is assured \cite{Dona2018, article:Dona_etal_2019_numerical_study_lorentzian_EPRL_spinfoam_amplitude}, it is not possible to have a unique prescription to set the optimal $\Delta l$ to get an acceptable convergence, since it depends on the details of data such as the face spins $j_f$ and the Immirzi parameter. Furthermore, the convergence strongly depends on the structure of the 2-complex, and there is still no method that allows to estimate the error made in truncating the sum over the auxiliary spins. In section \ref{subsec:extrap_algorithm}, I describe a numerical property of the convergence that allows extrapolating the limit $\Delta l \rightarrow \infty$ for the self-energy. 
\subsection{EPRL self-energy amplitude} 
\label{subsec:EPRL_self_energy}
According the procedure described above, starting from the BF amplitude \eqref{eq:BF_intertwiners_amplitude}, I attach a booster function between the bulk intertwiners, choosing the boundary intertwiners $i \equiv (i_{+}. i_{-})$ as the gauge fixed ones in the two vertex amplitudes which form the self-energy spinfoam, which is the most convenient choice from a computational point of view. Here I consider the case \cite{MY_PAPER} in which the two boundary intertwiners have the same value. When this happens, the fully algebraic compact expression of amplitude becomes\footnote{in the following omit the EPRL subscript for the amplitude, since every time I write $W$ (without the BF subscript) I always implicitly refer to the EPRL model}:
\begin{align}
\label{eq:EPRL_self_energy_amplitude_compact}
W_{} \left( j, i, \gamma; K , \Delta l \right) = &  \sum\limits_{j_{ab}}^{K}\prod\limits_{(a,b)}(2 j_{ab} +1) 
\sum\limits_{i_e}
A_{v}(j, j_{ab}, i, i_e, \gamma ; \Delta l)^2
\end{align}
where $q = 2,...5$, and the EPRL vertex amplitude is defined in \eqref{eq:vertex_amplitude}. The EPRL self-energy amplitude \eqref{eq:EPRL_self_energy_amplitude_compact} is the main object of our analysis. The dependence on the Immirzi parameter $\gamma$ is ``hidden" inside the booster functions, which appear in the definition of the single vertex amplitude \eqref{eq:vertex_amplitude}. In the EPRL self-energy amplitude \eqref{eq:EPRL_self_energy_amplitude_compact} I explicitly emphasized the \textit{artificial} dependence on the two cut-off $K$ and $\Delta l$. 
I shall always consider the two boundary intertinwers $i \equiv (i_{+}. i_{-})$ to have the same value. When this is the case, the asymptotic divergence scaling remains essentially\footnote{a part from irrelevant numerical fluctuations} unchanged by modifying the value of the intertwiners. Since it is necessary to choose a specific value of $i$ for the computations, in the following I use $i = (0,0)$ by default. This completes the derivation of amplitude we considered in \cite{MY_PAPER}, and now I describe the numerical results obtained.
\section{Divergence analysis} 
\label{sec:Div_analysis}
\begin{figure}[tb]
\centering
\includegraphics[width=7cm]{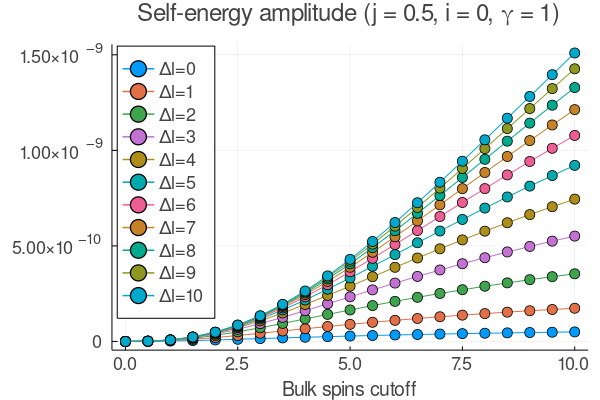}
\caption{\label{fig:pentachorscaling_cutoff10} \emph{Divergence of the EPRL self-energy amplitude \eqref{eq:EPRL_self_energy_amplitude_compact} computed numerically. All boundary spins $j \equiv j_f$, where $f=1...4$, are equal to $\frac{1}{2}$, while boundary intertwiners $i_{+}, i_{-}$ are both set to zero. The plots with $i_{+}, i_{-}$ equal to $1$ are identical.}}
\end{figure}

The amplitude \eqref{eq:EPRL_self_energy_amplitude_compact} has been studied numerically. Recalling that the analysis must be limited to a small number of parameters' configurations, the goal of the latter consists in answering the following question (for further numerical studies I refer again to \cite{MY_PAPER}, which addresses more questions), for which there are currently no analytical methods of investigation: what is the exact asymptotic scaling of the EPRL divergence, and is there a dependence on the Immirzi parameter? \\
The logical step is to proceed in the same way as we did with the BF amplitude \eqref{eq:BF_intertwiners_amplitude}, that is, fix the boundary state entirely to the lowest value $j = 0.5$, and compute the amplitude as a function of the cut-off $K$ on the internal faces. It is reasonable to do so using several different values of the Immirzi parameter $\gamma$. However, compared to the purely $SU(2)$ case, the EPRL amplitude $\eqref{eq:EPRL_self_energy_amplitude_compact}$ has an additional cut-off $\Delta l$ necessary to truncate the sum on the auxiliary spins $l_f$ in the single vertex amplitudes \eqref{eq:vertex_amplitude} which constitute self-energy. This fact makes the EPRL analysis much more complicated due to the reasons explained in section \ref{sec:From_BF_to_EPRL}. In section \ref{subsec:extrap_algorithm} I present a property of the convergence which allows to extrapolate the limit $\Delta l \rightarrow \infty$. Since for computations it is necessary to select a specific value of the Immirzi parameter, I first illustrate the divergence scaling extrapolation algorithm by selecting a specific value for the latter. The analysis for different $\gamma$ values is similar.  \\ \\
The result of the computation of the amplitude \eqref{eq:EPRL_self_energy_amplitude_compact}, for increasing values of $\Delta l$ and $K$, is shown in Figure \ref{fig:pentachorscaling_cutoff10}. 
From a simple qualitative analysis we infer that the convergence is faster for a reduced bulk spins cutoff, while it becomes slower as $K$ increases. Strictly speaking, the numerically computed curves turn out to be a \textit{lower bound} to the EPRL divergence, since the latter is recovered only in the limit $ \Delta l \rightarrow \infty $. Since further increasing numerically the parameter $\Delta l$ requires exponentially increasing the computation times, it is necessary to introduce an algorithm for extrapolating the above limit. 
\subsection{Extrapolation algorithm for the amplitude}
\label{subsec:extrap_algorithm}
\begin{figure}[tb]
\centering
\includegraphics[width=7cm]{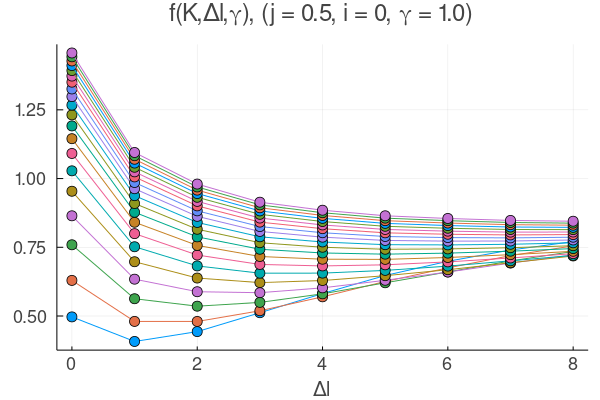}
\caption{\label{fig:Diff_and_ratios} \emph{Plot of the function $f(K, \Delta l, \gamma)$ defined as in \eqref{eq:f_function}. I plot the values of $K$ between $\frac{1}{2}$ and $10$ in ascending order from bottom to top.}}
\end{figure}

In order to derive the full EPRL amplitude, we study the convergence of \eqref{eq:EPRL_self_energy_amplitude_compact} in the parameter $\Delta l$ based on the data of Figure \ref{fig:pentachorscaling_cutoff10}, and we extrapolate infinite shell limit based on this trend. In order to do so, we plot the ratios of the differences between adjacent curves of Figure \ref{fig:pentachorscaling_cutoff10}, at fixed $K$, as a function of $\Delta l$. That is, we study the function:
\begin{equation}
\label{eq:f_function}
%\sum\limits_{\Delta l = 0}^{\infty} 
f(K, \Delta l, \gamma) \equiv \frac{W(K, \Delta l + 2, \gamma) - W(K, \Delta l + 1, \gamma)}{W(K, \Delta l + 1, \gamma) - W(K, \Delta l, \gamma)}.    
\end{equation}
The result is shown in Figure \ref{fig:Diff_and_ratios}. 
Numerical data shows that the convergence of \eqref{eq:EPRL_self_energy_amplitude_compact} in the shell parameter $\Delta l$, for each bulk spins cut-off $K$, is such that the function \eqref{eq:f_function} is first a decreasing function of $\Delta l$, then it remains constant in the convergence phase. Therefore, it exists an integer $N$ such that:
\begin{equation}
\label{eq:pipeline_property}
W(K, \Delta l + 1, \gamma) - W(K, \Delta l, \gamma) \approx \left( c_{K, \gamma} \right)^{\Delta l}  \hspace{4mm} \text{for} \hspace{2mm} \Delta l \geq N
\end{equation}
where the coefficient $c_{K, \gamma}$ is roughly equal to function \eqref{eq:f_function} for the highest number of shells computed numerically. So, if the amplitude has been calculated up to $N$ shells, we approximate the coefficient $c_{K, \gamma}$ as:
\begin{equation}
\label{eq:c_coefficient}
c_{K,\gamma} \equiv f\left(K, N-2, \gamma  \right).
\end{equation}
For the extrapolation of the EPRL amplitude we used $N = 10$, 
even if Figure \ref{fig:Diff_and_ratios} shows that property \eqref{eq:pipeline_property} becomes evident even for lower values. 
Once the convergence is reached, that is, when $\frac{W(K, \Delta l+1, \gamma)}{W(K, \Delta l, \gamma)} \approx 1$, function \eqref{eq:f_function} shows some slight fluctuations by further increasing the shells\footnote{as we shall see, this fluctuations are not relevant}. This is evident for low bulk spins cut-off values, where convergence in $\Delta l$ is extremely fast, as shown in Figure \ref{fig:Diff_and_ratios}. Property \eqref{eq:pipeline_property} allows to obtain a good estimate of the EPRL curve by using the equation:
\begin{equation}
W(K, \gamma) \equiv \lim_{\Delta l \to \infty} W(K, \Delta l, \gamma) \approx W(K, N-1, \gamma) + \frac{W(K, N, \gamma) - W(K, N-1, \gamma)}{1 - c_{K, \gamma}}
\label{eq:extrapolation_equation}
\end{equation}
Equation \eqref{eq:extrapolation_equation} is obtained by using the known limit of the geometric series:
\begin{equation}
\sum\limits_{l = N}^{\infty} \left( c_{K, \gamma} \right)^{l} = \frac{\left( c_{K, \gamma} \right)^N}{1 - c_{K, \gamma}} \approx \frac{W(K, N, \gamma) - W(K, N-1, \gamma)}{1 - c_{K, \gamma}}
\end{equation}
where in the second passage we used the property \eqref{eq:pipeline_property}. In Figure \ref{fig:Combined_scaling_label} we plot the EPRL amplitude extrapolated with equation \eqref{eq:extrapolation_equation} along with the curves in Figure \ref{fig:pentachorscaling_cutoff10}. \\ 

\begin{figure}[tb]
\centering
\includegraphics[width=7cm]{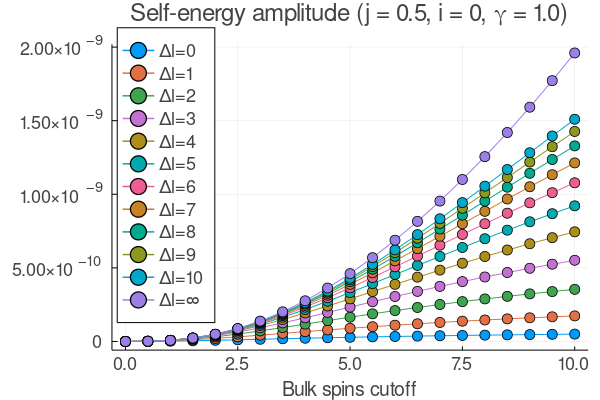}
\caption{\label{fig:Combined_scaling_label} \emph{EPRL asymptotic scaling curve extrapolated with the equation \eqref{eq:extrapolation_equation}, plotted together with those computed numerically, that is, the curves of Figure \ref{fig:pentachorscaling_cutoff10}. Notice that for $K \leq 4$, where the convergence is reached numerically, the extrapolated point essentially coincides with the last computed one. }}
\end{figure}
Let's discuss the above extrapolation scheme:
\begin{itemize}
    \item Extrapolating the limit $\Delta l \rightarrow \infty$ of \eqref{eq:EPRL_self_energy_amplitude_compact} for low values of the bulk spins cut-off $K$, for which a good convergence in the auxiliary spin sum has already been reached numerically, equation \eqref{eq:extrapolation_equation} provides a value which essentially coincides with the last computed point. This happens because, despite the fluctuations of function \eqref{eq:f_function} for low $K$, the difference $W(K, N, \gamma) - W(K, N-1, \gamma)$ is extremely small. Therefore, the only relevant contribution to the extrapolated amplitude comes from the computed one, which is a good approximation of the EPRL model.
    \item For each bulk spins cut-off value $K$, the corresponding EPRL amplitude is extrapolated \textit{independently} from the other values the latter. By doing so, we actually see that the resulting amplitude can be approximated extremely well by a polynomial fit $W(K,\gamma) = a + b K^{c}$ as a function of the bulk spins cut-off $K$.
    \item The specific properties of the convergence in the auxiliary spins sum of the single vertex amplitude \eqref{eq:vertex_amplitude} are still unexplored. It is possible that property \eqref{eq:pipeline_property} also manifests for spinfoam amplitudes defined on a more elaborate 2-complex with respect to the triangulation described in section \ref{sec:self_energy_diagram}. If this would be the case, the above extrapolation scheme could allow to obtain a good estimate of the EPRL model by using a negligible amount of computational resources compared to that necessary to reach a good approximation by using solely numerical techniques.
    \item In order to apply equation \eqref{eq:extrapolation_equation} it is necessary to know $W(K, N, \gamma)$, $W(K, N-1, \gamma)$, $W(K, N-2, \gamma)$. That is, only three amplitudes must be computed numerically (for $N$ sufficiently high). Despite this, we still opted to compute all the amplitudes $W(K, \Delta l, \gamma)$ for $\Delta l = 0,1...N$. This was done both to test property \eqref{eq:pipeline_property} and, on the other hand, to estimate the effectiveness of the extrapolation with a qualitative comparison between all the numerically calculated amplitudes and the extrapolated one. Furthermore, since the curves in Figure \ref{fig:Diff_and_ratios} are not exactly constant, that is, they start to increase when $\frac{W(K, \Delta l+1, \gamma)}{W(K, \Delta l, \gamma)} \approx 1$, the extrapolated curve is an approximation, which improves by increasing $N$ in equation \eqref{eq:extrapolation_equation}. Therefore, the higher is the number of shells computed numerically, the better is the approximation represented by the extrapolated amplitude. In \cite{MY_PAPER} we discuss this in more detail and we also provide further numerical confirmation of the effectiveness of this extrapolation algorithm.
\end{itemize}
\subsection{Scaling of the divergence and role of the Immirzi parameter}
\label{subsec:scaling_and_Immirzi}
Using the above scheme in \cite{MY_PAPER}, we study the scaling of the asymptotic divergence of the EPRL self-energy amplitude for $9$ different values of the Immirzi parameter. We choose an approximately uniform sampling of $\gamma$ between $0.1$ and $10$, arguing that a significant effect of the latter in the scaling should occur in a range which spans two orders of magnitude. We fit the curves with a function:
\begin{equation}
W(K,\gamma) = a + b K^{c}
\end{equation}
where $a,b,c$ are real coefficients. The values\footnote{recall that we neglected all the dimensional factors which, in the amplitude \eqref{eq:EPRL_self_energy_amplitude_compact}, are constant as a function of $K$} of $W(K,\gamma)$ are reported in \cite{MY_PAPER}. %table \ref{tbl:coefficient_values}.
\begin{figure}[tb]
\centering
\includegraphics[width=4cm]{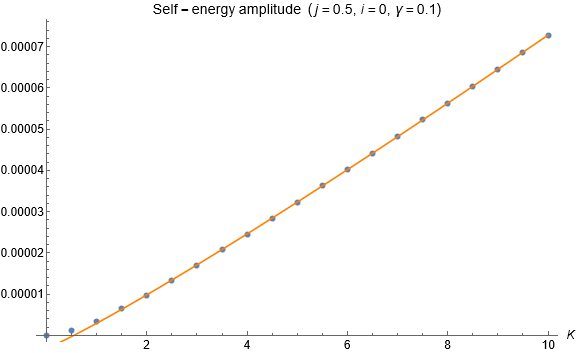}
\includegraphics[width=4cm]{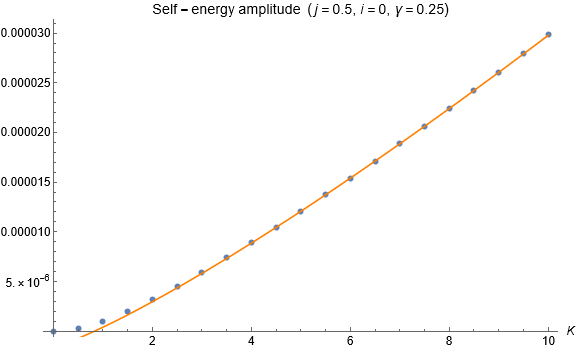}
\includegraphics[width=4cm]{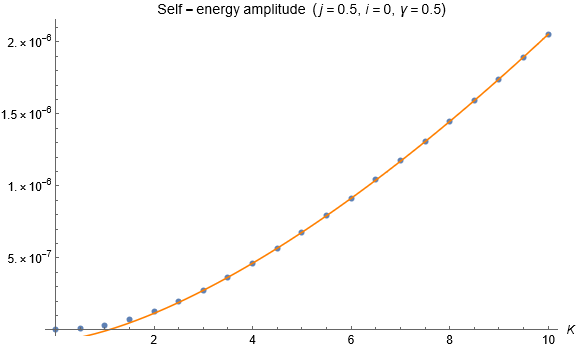}
\includegraphics[width=4cm]{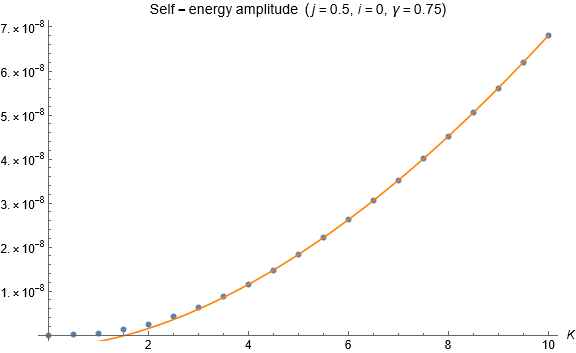}
\includegraphics[width=4cm]{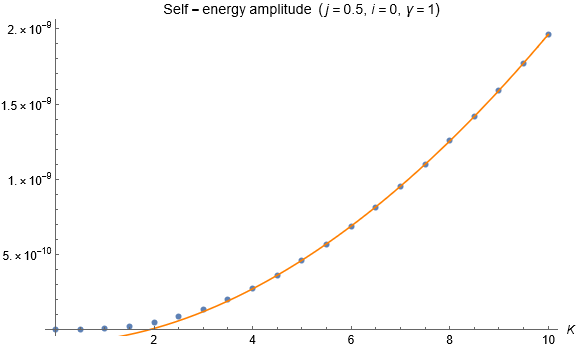}
\includegraphics[width=4cm]{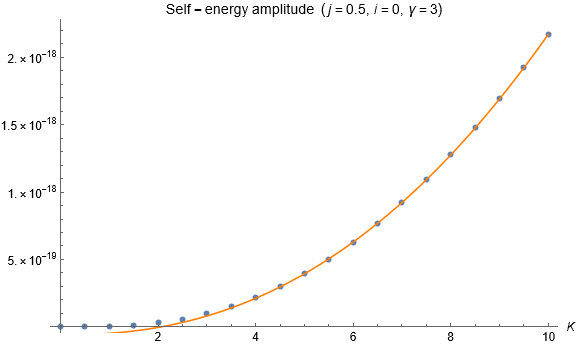}
\caption{\label{fig:Combined_scaling_label} \emph{EPRL asymptotic scaling curve extrapolated with the equation \eqref{eq:extrapolation_equation}, plotted together with those computed numerically, that is, the curves of Figure \ref{fig:pentachorscaling_cutoff10}. Notice that for $K \leq 4$, where the convergence is reached numerically, the extrapolated point essentially coincides with the last computed one. }}
\end{figure}
%
%\begin{center}
%\begin{table}[ht]
%\caption{Fit coefficients table}
%\begin{tabular}{ |p{3cm}||p{3cm}|p{3cm}|p{3cm}|  }
% \hline
% \multicolumn{4}{|c|}{ $W(K,\gamma) = a + b K^{c}$ $ \hspace{2mm}  (j = 0.5, i = 0)$, $K \in [0, 10]$} \\
% \hline
%Immirzi parameter & $a$ & $b$ & $c$ \\
% \hline
% $\gamma = 0.1$  & $-3.0884*10^{-6}$    & $6.0304*10^{-6}$ &  1.1  \\
% $\gamma = 0.25$ & $-1.58614*10^{-6}$ & $1.9783*10^{-6}$ & 1.2 \\
% $\gamma = 0.5$ &  $-7.5523*10^{-8}$  &  $6.7306*10^{-8}$   & 1.5 \\
% $\gamma = 0.75$ & $-1.7153*10^{-9}$ & $1.10779*10^{-9}$  & 1.8 \\
%$\gamma = 1$ & $-8.6894*10^{-11}$ & $2.5801*10^{-11}$ & 1.9 \\
%$\gamma = 3$   & $-6.0026*10^{-20}$ & $1.1186*10^{-20}$ & 2.3  \\
% $\gamma = 5$   & $-4.5405*10^{-25}$ & $1.0642*10^{-25}$ & 2.3  \\
% $\gamma = 7.5$   & $-2.35075*10^{-29}$ & $8.25374*10^{-30}$ & 2.3  \\ 
%$\gamma = 10$   & $-1.7872*10^{-32}$ & $9.2770*10^{-33}$ & 2.3  \\
% \hline
%\end{tabular}
%\label{tbl:coefficient_values}
%\end{table}
%\end{center}
%
Figure \ref{fig:Combined_scaling_label} shows that the amplitude is strongly suppressed as $\gamma$ increases. Moreover, the divergence turns out to be well fitted by a linear scaling when the Immirzi parameter is small enough, while for $\gamma \geq 1$ the curve is approximated by a quadratic function in the range $K \in \left[0,10 \right]$. Unfortunately, the huge computation cost required by increasing the bulk spins cut-off in the divergence analysis prevents us from testing the divergence for $K > 10$. It is possible that the divergence scaling is actually independent of the Immirzi parameter, and that the different asymptotic trends, as $\gamma$ varies, are only an effect of the fact that the range of the bulk spins cut-off $K$ is too small. On the other hand, it is also possible that the Immirzi parameter plays an effective role in modifying the asymptotic scaling of self-energy amplitude. This deserves future investigations. \\
In any case, the numerically observed scaling in the range $K \in \left[0,10 \right]$, with boundary parameters $j = 0.5, i = 0$ falls within the upper and lower bounds present in the literature. In particular, unlike the BF self-energy divergence (see equation \eqref{fig:bf_scaling}), the role of the destructive interference between the oscillations of the booster functions $B_4 \left( j_f, l_f ;  i ,  k \right)$ and the $\{15j \}$ symbols in the sum over the bulk spins $j_{ab}$ implies that the divergence is considerably dumped. This was not expected, since it has been shown that in the three-dimensional EPRL model, the upper bound provided by the algorithm proposed in \cite{Dona:2018infrared} provides an excellent estimate of the divergence.
\section{Boundary observables} 
\label{sec:Boundary_observables}
In this section we compute some spinfoam boundary observables. We focus on the normalized dynamical expectation value of geometrical operators, that is: 
\begin{equation}
\label{eq:dynamic_exp_value}
\langle O \rangle = \frac{ \langle W | O | \Psi \rangle }{ \langle W | \Psi \rangle }
\end{equation}
where the bra $W$ contains the propagator, namely the dynamics, while the ket $\Psi$ turns out to be the tensor product of the \textit{in} and \textit{out} states of the LQG Hilbert space. With the term ``propagator" we refer to the square matrices (they are such since the self-energy triangulation has two boundary tetrahedra) in which the element $a, b$ corresponds to the EPRL self-energy amplitude with $a = i_+, b = i_{-}$. The observable $O$ therefore contains the dynamic correlations. \\ 
The booster function $B_4 \left( j_f, l_f ;  i ,  k \right)$ are interpreted as a quantum tetrahedron being boosted among adjacent frames: the two sets $j_f$ and $l_f$ describe the four areas of the tetrahedron in the two frames connected by a boost, and the two intertwiners $i$ and $k$ describe the quantum intrinsic shape of the tetrahedron \cite{dona2020}. From a numerical point of view it is an excellent check, especially considering that numerical calculations carried out with the EPRL model are still in their primordial stages \cite{Dona2019, Dona:2018infrared, article:Dona_etal_2019_numerical_study_lorentzian_EPRL_spinfoam_amplitude}, and, as far as we know, so far there are no such numerical computations (ie dynamic expectation value of geometric operators) with divergent spinfoams. \\
In \cite{MY_PAPER} we find that the boundary observables are not affected in any way by the presence of the spinfoam divergence in the bulk. Furthermore, while in the divergence analysis we see an important contribution due to the number of shells $\Delta l$ of the amplitude \eqref{eq:EPRL_self_energy_amplitude_compact}, for the observables there is no trace of a dependence in that sense\footnote{at least for the ones that we computed}. In fact, an excellent approximation of the correct geometric value is obtained with the approximation $\Delta l = 0$. The fact that the number of shells is completely irrelevant in the computation of geometric boundary observables allows them to be computed very fast.
\subsection{Angles}
\label{subsec:angles}
\begin{figure}[tb]
    \centering   
        \includegraphics[width=4cm]{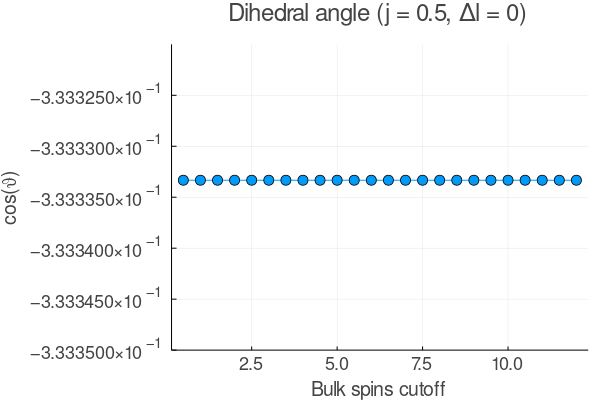}
        \includegraphics[width=4cm]{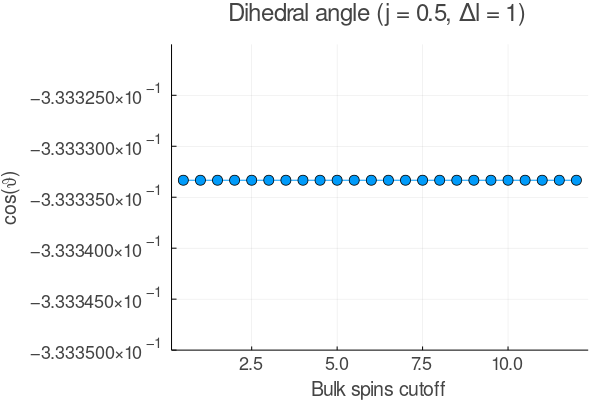}
        \includegraphics[width=4cm]{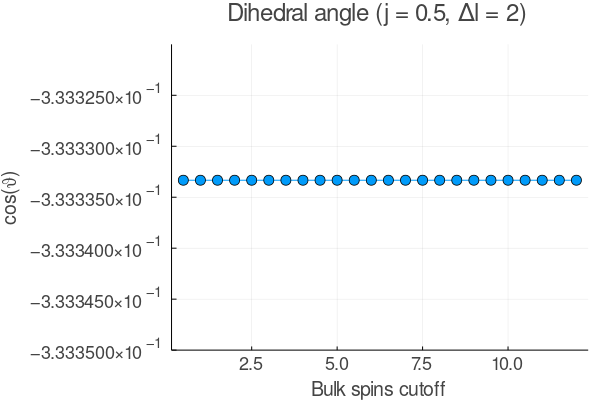}
        \includegraphics[width=4cm]{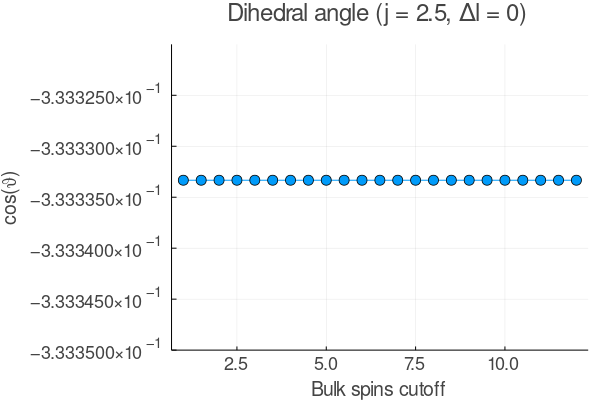}
        \includegraphics[width=4cm]{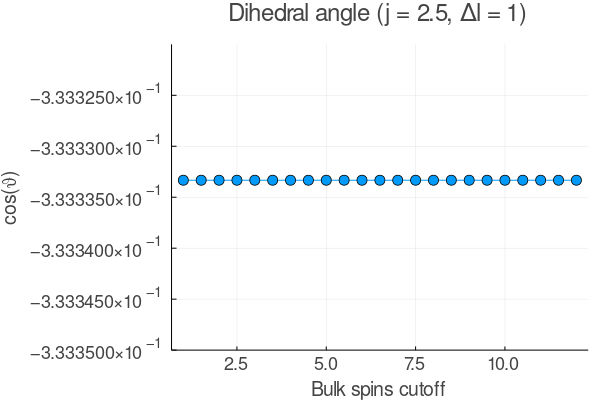}
        \includegraphics[width=4cm]{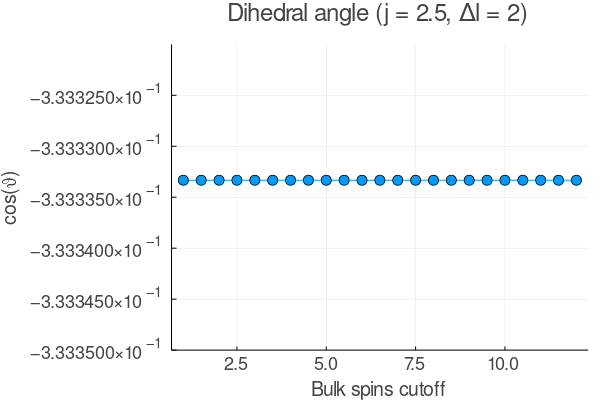}
    \caption{\label{fig:angles} \emph{Dynamic expectation value of the cosine of the external dihedral angle operator \eqref{eq:dynamic_angle}. We plot the angle for different values of $\Delta l$, showing that an excellent agreement with the value $-0.\Bar{3}$ is obtained with the approximation $\Delta l = 0$. We used $\gamma = 0.1$. }}
    \end{figure}

The shape of the tetrahedra in twisted geometry is measured by the angle operator:
\begin{equation}
  \label{eq:geom-angleop}
  A_{ab} |i_{\pm} \rangle = \cos(\theta_{ab}) |i_{\pm} \rangle
\end{equation}
which is interpreted as the cosine of the external dihedral angle between the faces $a$ and $b$ of the tetrahedron defined on the nodes $\pm$. The spinnetwork basis states diagonalize the dihedral angle $\theta_{ab}$ between faces $a$ and $b$. The equation for measuring the dihedral angle $\cos(\theta_{ab})$ of $|i_{\pm} \rangle$ in terms of intertwiner spin $i_{\pm}$ was derived in \cite{Gozzini_primordial} and it reads:
\begin{equation}
  \label{eq:geom-angleformula}
  \cos(\theta_{ab}) = \frac{i_+(i_+ +1) - j_a(j_a+1) - j_b(j_b+1)}{2\sqrt{j_a(j_a+1)j_b(j_b+1)}}.
\end{equation}
We consider the expectation value \eqref{eq:dynamic_exp_value} of the angle operator \eqref{eq:geom-angleop} in any of the two (equal) boundary regular tetrahedra of the triangulation using the spinnetwork state. According to the recoupling basis $(j_1,j_2)$, we focus on the angle between faces $1$ and $2$. The expectation value can be computed as: 
\begin{equation}
\frac{\langle W | A_{12} | W \rangle}{ \langle W  | W \rangle } = \frac{ \sum\limits_{i_{\pm}} \left[ W(j, i, K) \right]^2 \cos(\theta_{12})}{ \sum\limits_{i_{\pm}} \left[ W(j, i, K)  \right]^2}.
\label{eq:dynamic_angle}
\end{equation}
Carrying out the numerical computation of \eqref{eq:dynamic_angle}, we obtain a value that is in agreement with the geometric value of the external angle of a regular tetrahedron up to the tenth significant digit, as shown in Figure \ref{fig:angles}.
This is consistent with the fact that we are looking at the only angle which is completely sharp, while the others turn out to be spread. In fact, according to the Heisenberg uncertainty principle, since different angle operators do not commute, only one of the dihedral angles can be determined.
We also analyzed the boundary states of the self-energy amplitude by using the Livine-Speziale coherent intertwiners, which allow to define a superposition of spin-network states peaked on a classical geometry with minimal spread. A wave packet peaked on a classical triangulated geometry, that can be viewed as a coherent state in the Hilbert space \eqref{eq:Hilbert_space}, is obtained by combining coherent intertwiners at each node. In this case, all the dihedral angles are minimally spread around the classical values, and this is exactly what we verify numerically. The plots obtained are qualitatively identical to those in Figure \ref{fig:angles}, except that the computed value of the angle is approximately\footnote{up to the first $2$ significant digits} that of the external dihedral angle of a regular tetrahedron. This is due to the fact that the square modulus of the coefficients of the coherent states, in the spin-intertwiner basis, turns out to be a a distribution centered around the value of the intertwiner which determines the semiclassical value of the dihedral angle, and the width of the distribution is much higher as the spins are low. For the computation of the volumes, we refer to \cite{MY_PAPER}. 

\section{Conclusions}
\label{sec:conclusions}
I have presented the application \cite{MY_PAPER} of new computational techniques applied to the study of the infrared divergence represented by the self-energy EPRL amplitude. The divergence scaling obtained falls within the upper and lower bounds in the literature. Particular emphasis was placed on the role of the Immirzi parameter in the asymptotic divergence. I presented the extrapolation method introduced to overcome the computational cost represented by the convergence in the shell parameter $\Delta l$, which can potentially be used in other contexts.
The analysis of the boundary angles shows that the latter are finite and consistent with classical geometry, despite the divergence in the bulk, and that the approximation represented by the shells does not play any relevant role in the dynamical expectation values of the latter. \\
We thank all the collaborators of this work, in particular Francesco Gozzini, without whom this would not have been possible, as well as Carlo Rovelli for useful discussions and comments. I explicitly thank my supervisor Francesca Vidotto for the constant support and for an accurate review in the drafting of this paper. I also thank Pietro Dona for countless contributions in various stages of \cite{MY_PAPER}. Finally, we also thanks the Compute Canada staff for their help in using the Graham and Cedar clusters. This work is supported by the NSERC Discovery Grant.

\bibliographystyle{ws-procs961x669}
\bibliography{ws-pro-sample}

\end{document}